# Laser optomechanics

Weijian Yang[1], S. Adair Gerke[1], Kar Wei Ng[1], Yi Rao[2], Christopher Chase[2] & Connie J. Chang-Hasnain[1]

**Cavity optomechanics[1-5] explores the coupling between the optical field and the mechanical oscillation to induce cooling[6-10] and regenerative oscillation[11-16] in a mechanical oscillator. So far, optomechanics relies on the detuning between the cavity and an external pump laser, where the laser acts only as a power supply. Here, we report a new scheme with mutual coupling between a mechanical oscillator that supports a mirror of a vertical-cavity surface-emitting laser (VCSEL) and the optical field, greatly enhancing the light-matter energy transfer. In this work, we used an ultra-light-weight (130 pg) high-contrast-grating (HCG) mirror in a VCSEL, whose reflectivity spectrum is designed to facilitate strong optomechanical coupling, to demonstrate optomechanically-induced regenerative oscillation of the laser optomechanical cavity with > 550 nm self-oscillation amplitude of the micro-mechanical oscillator, two to three orders of magnitude larger than typical[11-16]. This new scheme not only offers an efficient approach for high-speed wavelength-swept sources, but also has far-reaching significance in the realization of quantum entanglement of macroscopic objects[17] and ultrasensitive measurement of displacements and forces[18].**

The coupling of an optical field and a mechanical oscillator through forces induced by light has been a subject of long-standing interest and an important tool to study fundamental physics[1-5]. Numerous recent experiments have used light-induced forces in optical cavities to demonstrate laser cooling[6-10] and regenerative oscillation[11-16]. The prevalent experiment apparatus consists of a passive optical cavity housing a mechanical oscillator with a population of circulating photons produced by an external low-noise continuous-wave laser. In such setups, light can induce forces on a mechanical resonator directly through radiation pressure or indirectly through photothermal forces, which result from thermal expansion of the structure under light-induced heating. While the optomechanical coupling can be adjusted by the frequency detuning between the cavity and the laser, the power coupled between the optical field and the mechanical oscillation mode is ultimately limited. Large mechanical oscillation strength implies a large detuning, but large detuning reduces the photon population and thus the optomechanical forces. Despite recent progress, regenerative mechanical oscillation has been limited to nanometer-scale amplitudes[11-16] for oscillators using radiation pressure and limited to low-kHz-scale frequencies for oscillators using photothermal effects[19]. Here, we show a new configuration that integrates the mechanical oscillator as a mirror in a laser cavity, as shown in Fig. 1, producing both high-speed and large-amplitude regenerative oscillation.

[1] Department of Electrical Engineering and Computer Sciences, University of California, Berkeley, CA 94720, USA
[2] Bandwidth10, Inc., San Jose, CA 94063, USA

The difference from conventional cavity optomechanics is that the cavity is embedded with the optical gain, serving as the laser itself to drive the mechanical oscillator and thus integrating an optomechanical system in a single device.

Collocating a VCSEL cavity and the mechanical oscillator has multiple functions. Firstly, the mechanical oscillation of the mirror results in periodic oscillations of the laser wavelength $\lambda$ and output power $P$, as shown in Fig. 1. Since the VCSEL cavity can be designed to be on resonance for a single wavelength over a wide range of wavelength, a unique attribute of a VCSEL, the optomechanical forces, which include both radiation pressure and photothermal forces, can stay enhanced over a broad wavelength range. Secondly, the negative damping induced by optomechanical forces is greatly strengthened by the delayed response of the optomechanical forces to the mechanical movement of the mirror, which arises from the long lifetime of the excited states in the gain medium. Furthermore, the wavelength and mechanical displacement dependence of mirror reflectivity $R$ and laser gain can be engineered independently in the ultra-light HCG, which could lead to interesting dynamics, such as extremely large oscillation amplitude and laser cooling. Self-oscillation using this scheme can thus be implemented in any mechanically-tunable laser possessing a mirror with wavelength-dependent reflectivity, a tuning mechanism with high Q-factor, and a large delay resulting from a sufficiently slow dynamic response to modulation. The new scheme facilitates a strong power coupling between the optical radiation field and the mechanical oscillator. We term it laser optomechanics.

An optomechanical laser naturally acts as a wide-range high-speed self-wavelength-swept light source. This has been long sought for high-speed energy-efficient 3D imaging, such as optical coherence tomography[20] and light detection and ranging[21]. Laser optomechanics immediately can address the demand for improved light sources in these fields. Additionally, replacing the external high-power laser source with a low-power, microscale VCSEL can open existing passive optomechanical technologies, such as clock generation, to portable applications.

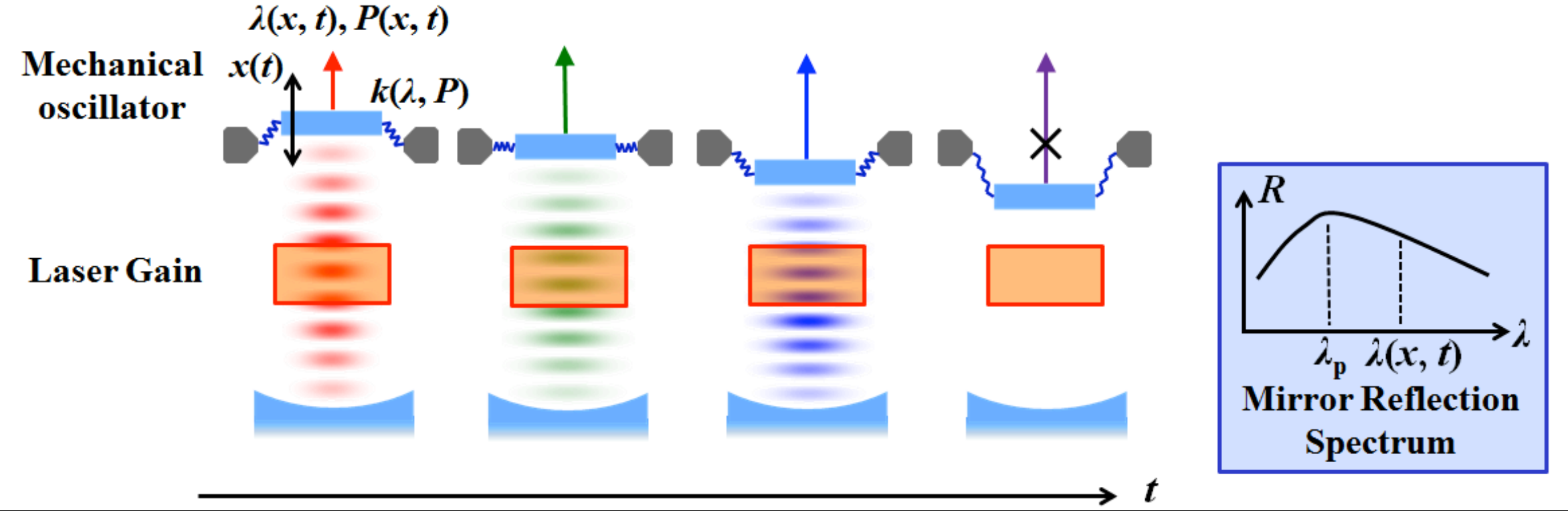

**Figure 1 | Schematic of experimental apparatus of laser optomechanics.** The optical cavity houses the laser source as well as a mechanically moveable mirror with a reflection spectrum $R(\lambda)$. The laser can be either optical pumped or electrically pumped. As the mirror oscillates with displacement $x(t)$, the laser

output wavelength $\lambda(x, t)$ and power $P(\lambda, t)$ changes with respect to time $t$. The spring constant $k$ is a function of the optical wavelength and laser output power, which illustrate the optical spring effect. The various images visualize this wavelength changes: the lasing wavelength changes from red to blue as the effective cavity length shortens due to the mirror oscillation. Depending on the mirror reflectivity and gain, the mechanical oscillation can be sustained over a large displacement range.

The optomechanical laser cavity demonstrated here uses an ultra-light weight (130 pg) high-contrast-grating (HCG) vertical-cavity surface-emitting laser (VCSEL)[22-24] electrically-pumped lasing at 1550 nm. The HCG is an InP grating with near-wavelength period, designed to reflect >99.5% surface-normal light and serve as the top mirror of the VCSEL[25-26]. The HCG platform offers substantial design flexibility. It can be designed such that the high reflection is selected for light polarized in parallel or in perpendicular to the grating bar. The former is called TE HCG, while the latter TM HCG. The optimal dimensions are quite different, particularly in thickness. For 1550 nm operation, the TE HCG is typically ~195 nm thick and the TM HCG is ~430 nm thick. In this paper, we show that lasers of both TE and TM designs exhibit laser optomechanical self-oscillation. Detailed device designs can be found in Reference 24. Additionally, the ultra-small size and thin, long support arms of the HCG mirror allow its temperature to vary significantly from that of the bulk of the chip, enabling photothermal distortion forces. As a result, the optomechanical forces described in this work include contributions from both radiation pressure and photothermal forces. The rest of the VCSEL cavity contains an active region with multiple quantum well and a distributed Bragg reflector (DBR) as the bottom mirror. Current confinement is provided with a wide-aperture proton implant, ensuring a low beam divergence angle of < 5 degrees[27]. In ambient temperature and pressure, the HCG-VCSEL lases with a DC threshold current of 5 mA and emits 1.7 mW at 12 mA bias with a single, fundamental transverse mode. The lasing wavelength of the VCSEL can be tuned by electrostatically displacing the HCG with a voltage across the HCG layer, which changes the cavity length. Fig. 2a shows the schematics of the HCG VCSEL, and Fig. 2c shows the scanning microscope image (SEM) of a TE HCG. The HCG mirror is 20 μm x 20 μm in size and 195 nm thick, with a total weight of only ~130 pg, two orders of magnitude lighter than a standard DBR mirror. The mirror is held by a double bridge spring structure. Under vacuum, this micro-electro-mechanical (MEMS) structure exhibits a high mechanical quality factor $Q_m$, which can be measured by Laser Doppler vibrometry (LDV) (see Methods for details). Figure 2b shows LDV characterization of a fundamental-mode mechanical resonance of a TE-type device under small-signal electrostatic white noise actuation, fitted with a Lorentzian function to show a quality factor $Q_m$ = 3,640 at a frequency of 147 kHz. It is important to note that the VCSEL cavity is not pumped, and the wavelength of the external laser used in the LDV measurement is far from the VCSEL cavity resonance. Thus, LDV directly measures the mechanical property of the MEMS HCG unperturbed by the VCSEL's optomechanical effects.

The HCG self-oscillates at its mechanical resonance frequency with the VCSEL biased above threshold with 9 mA DC current in a vacuum of 2e-5 Torr. The self-oscillation of the HCG can be visualized by the SEM images of the HCG self-oscillating at fundamental mechanical mode, shown in Fig. 2d with 13 mA DC currents. A fast scanning rate is used so that stroboscopic effect can be visualized on the grating bars, from which the oscillation frequency and amplitude is calculated to be ~131 kHz and ~595 nm. To the best of our knowledge, this is the largest amplitude of regenerative oscillation in an optomechanical cavity. Compared with other MEMS-tunable lasers, the advantage of the HCG VCSEL is a combination of its light weight and high mechanical Q, which together reduce damping forces on the mirror, allowing optomechanical dynamics to more readily produce self-oscillation.

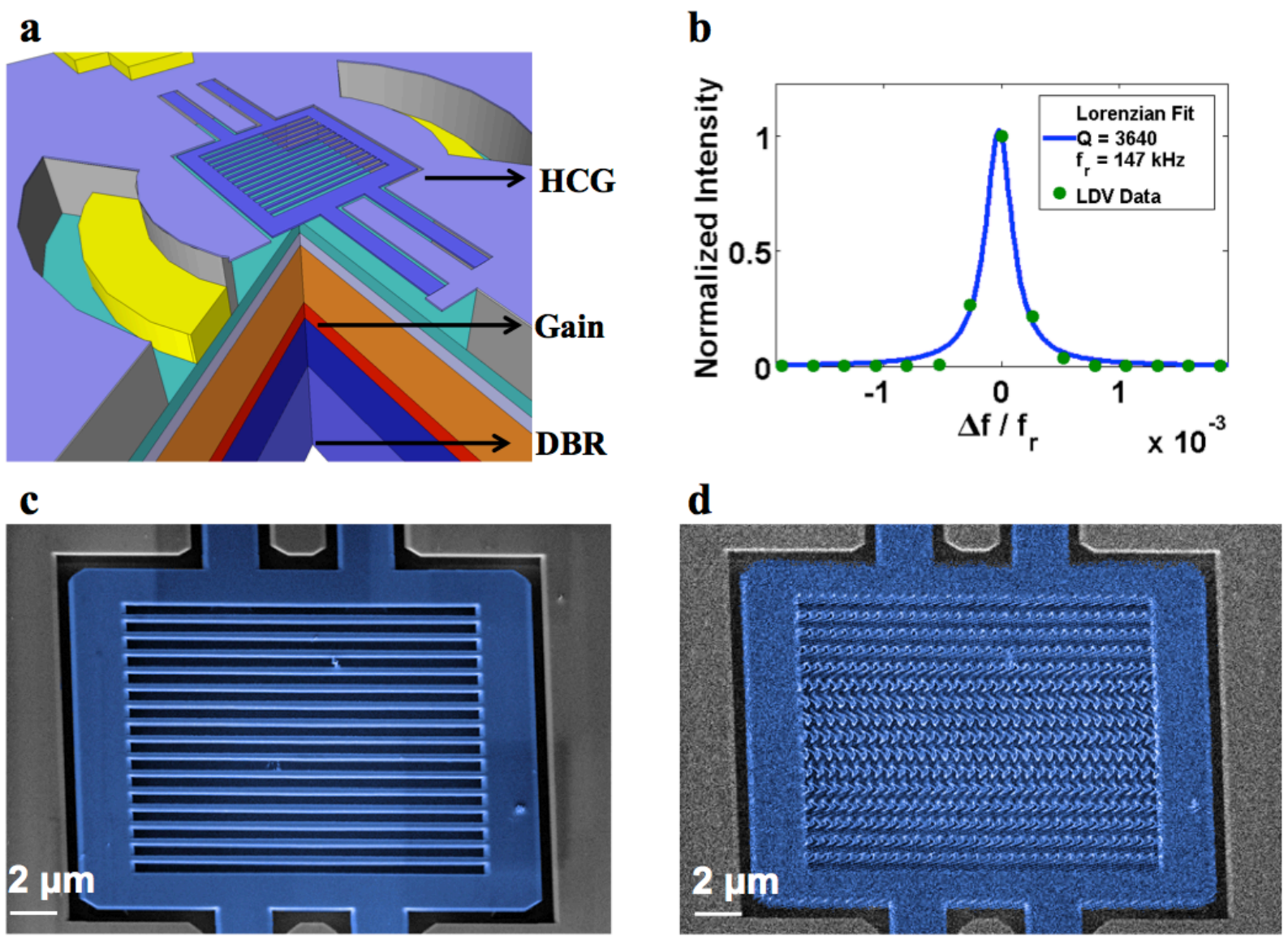

**Figure 2 | HCG VCSEL and the mechanical oscillation of HCG. a**, Schematic of the HCG VCSEL, which is composed of a capacitatively-actuated MEMS HCG top mirror, an active region with multiple quantum wells, and an immobile DBR bottom mirror. **b,** Laser Doppler vibrometry measurement of quality factor of fundamental mechanical mode of MEMS oscillation in vacuum, $Q_m = 3{,}640$. **c-d,** False color SEM image of the HCG. **c** shows the steady HCG when the bias current $I$ of the VCSEL is zero; **d** shows its oscillation at fundamental mode when $I = 13$ mA (laser threshold current is 5 mA at room temperature for this device). Due to the fast scanning rate of the SEM, the stroboscopic effect produces a periodic distortion of the HCG bars in the SEM image.

In laser optomechanics, the laser emission wavelength oscillates as the cavity mirror position oscillates. Figure 3a shows the results of time-resolved characterization for a typical VCSEL biased in mechanical oscillation state (see Methods for details). Simultaneously, the optical spectrum of the device was captured with an optical spectrum analyzer, as shown in Fig. 3b, confirming the total wavelength excursion of 23 nm results from the HCG oscillation with fundamental mechanical mode. Considering an intrinsic linewidth of 50 MHz for a typical HCG-VCSEL[23], the 23 nm wavelength shift corresponds to 57,500 times the intrinsic linewidth. Due to the longitudinal profile of the VCSEL mode, this wavelength shift translates into a ~560 nm HCG physical displacement, which is two to three orders of magnitude larger than those typically reported in optomechanical oscillators[11-16] .This large oscillation amplitude is a unique property attributed to laser optomechanics. The wavelength oscillation is accompanied by an oscillation of the optical power from the VCSEL. At the two ends of the oscillation excursion, the laser stops lasing. Thus the actual mechanical oscillation amplitude is larger. The power oscillation originates from the gain spectrum as well as the wavelength and mechanical deformation dependence on the mirror reflectivity. This translates into a change of the circulating optical power inside the cavity, and thus a temporal change of the optomechanical forces. When retarded by the dynamics of the laser cavity, the position-dependent optomechanical forces create instability the mechanical oscillator, leading to optomechanical self-oscillation. In large oscillation amplitude regime, there appear nonharmonic components in the motion. This can be seen in the asymmetry in the up-ramp and down-ramp of the wavelength excursion shown in Fig. 3a.

In conventional cavity optomechanics, optical input power and detuning are the two control knobs to tune the optomechanical dynamics[28]; in laser optomechanics, these correspond to the laser pumping strength and the cavity length, and specifically, the laser drive current and the tuning voltage of the HCG MEM structure. The laser current primarily changes the circulating power and the photon retardation, whereas the tuning voltage mainly modifies the spectrum overlap between the gain and cavity resonance, and thus the gradient of the radiation force. Fig 3c shows an example of this control. The optical wavelength swept range is measured with various laser currents and tuning voltages. With an increasing pump current, the gain boosts up, leading to an increased wavelength swept range. As the tuning voltage increases (cavity length shortens), the wavelength swept range decreases. This is most likely due to the laser experiencing a different radiation force gradient in the short wavelength regime. We will have a further discussion on the exact mechanism in the later section. Most importantly, this experiment shows that the opposite effect, mechanical cooling, can be realized in laser optomechanics as well with carefully engineering the laser parameters and the mechanical structure.

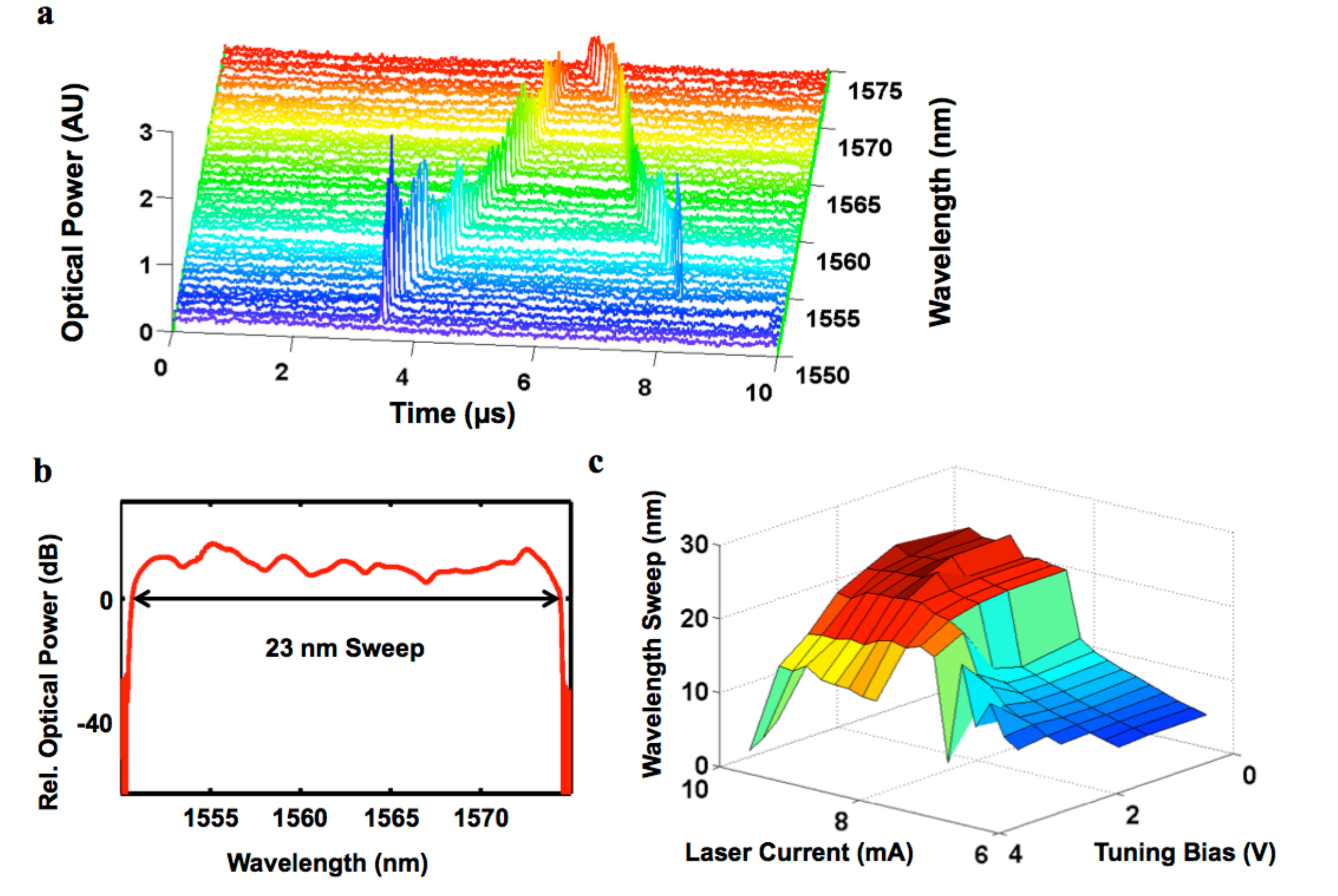

**Figure 3 | Regenerative oscillation of HCG VCSEL. a**, The time-resolved lasing wavelength of the HCG VCSEL with the HCG oscillating at fundamental mode. **b,** Optical spectrum obtained in an optical spectrum analyzer with high integration time, captured simultaneously with the time-resolved wavelength characterization of Fig. 3a, confirming the extent of the wavelength sweep. **c**, The wavelength swept range versus the laser current and tuning voltage. The laser current and tuning voltage can be used to tune the optomechanical dynamics.

In laser optomechanics, various mechanical modes can be excited. The mechanical damping factor and the gain from optomechanical forces determine the oscillation threshold. The modes with the lowest threshold will be first excited, just like that in a laser. Fig. 4 shows two different states of a VCSEL with a TM HCG: the pure fundamental mode at 345.6 kHz and a pure high order mode at 5.636 MHz. The TM HCG in this device has a size of 20 μm x 20 μm, and a thickness of 430 nm -- thicker than the TE HCG. The different states can be manipulated by the pump current and tuning voltage. When oscillating at high order mode, the 3 dB linewidth of the fundamental radio-frequency (RF) tone is 8 Hz, leading to a loaded mechanical Q of $7x10^5$. It is interesting to see there appear multiple RF tones (Fig. 4e). This could be attributed to the optical mode having a Gaussian spatial profile, which couples slightly differently for individual grating bars, leading to their mode splitting. The exact reason is under investigation. These interesting phenomena show the rich physics to be further explored in laser optomechanics.

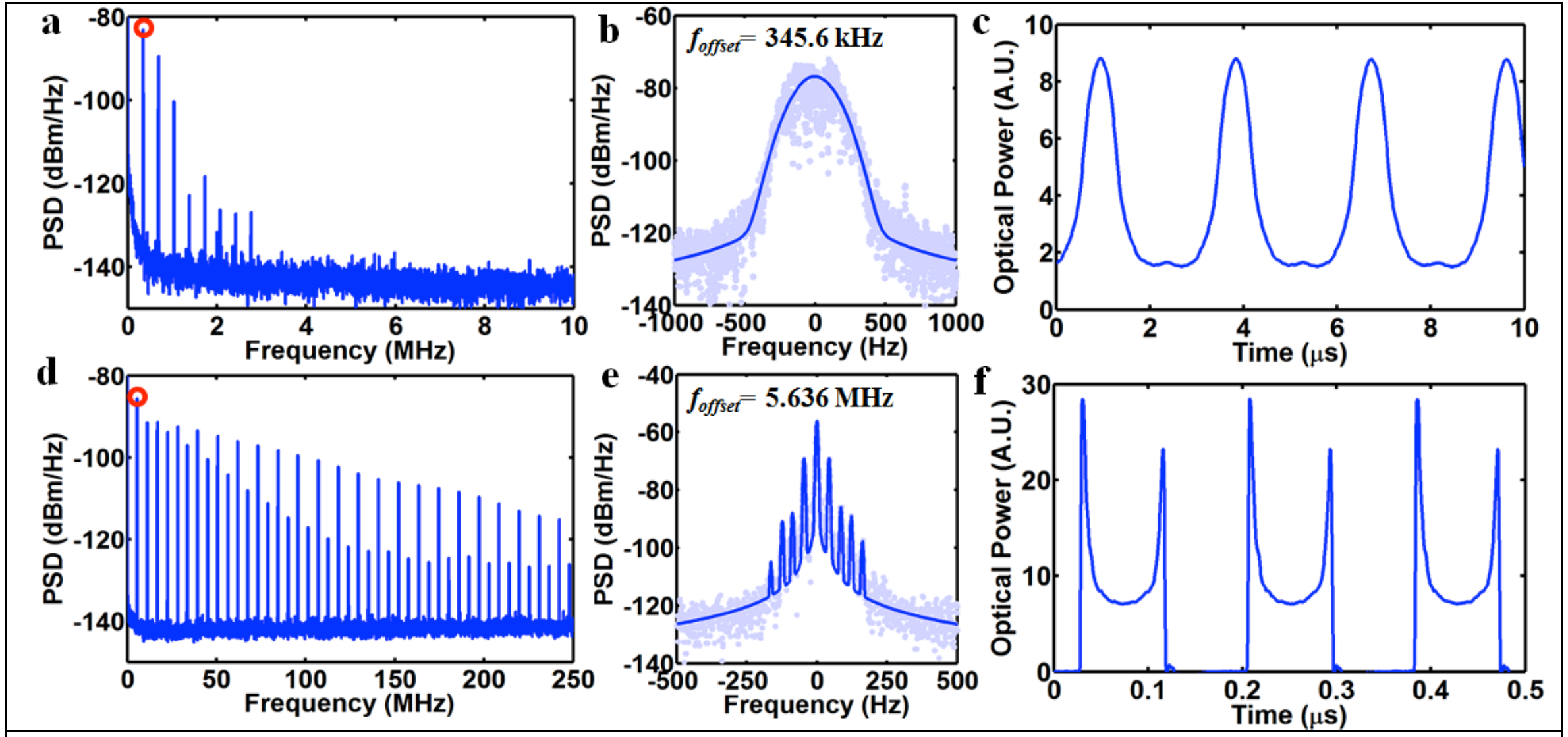

**Figure 4 | Regenerative oscillation of the HCG VCSEL in different mechanical oscillation modes. a-c,** Oscillation at fundamental mode; **d-f**, Oscillation at a high order mode. **a, d** shows the RF spectrum of the laser optical output power. **b**, **e** shows the zoom-in view of the first RF tone in **a**, **d** (marked with the red dot)**;** they are fitted with Voigt functions. The frequency offset in **b** and **e** is 345.6 kHz and 5.636 MHz respectively. **c, f** shows the laser optical output power in time-domain. The threshold current of the laser is 10.1 mA at room temperature. The laser current and tuning voltage in condition **a-c**, **d-f** is 19.2 mA, 10 V, 12 mA, 0 V respectively.

The mechanical modes of the HCG can be mainly divided into three categories, represented by Fig. 5a-c. Figure 5a depicts the fundamental mode, where the whole HCG moves in and out of its plane vertically. Figure 5b and 5c depict examples of the higher order mechanical modes. In general, we can assign a single generalized coordinate $x(t)$ to describe the displacement of the mirror. This oscillator is acted upon by the optomechanical forces $F_{OM}$, which comprises the radiation pressure force $F_{RP}$ and the photothermal force $F_{PT}$, the latter of which acts by expanding the grating bars or support arms of the HCG mirror under heating. For a mechanical oscillator with a mass of $m$ and a resonant angular frequency of $\omega_o$, the position $x(t)$ obeys a driven damped harmonic oscillation:

$$m\ddot{x}(t) + b\dot{x}(t) + m\omega_o^2 x(t) = F_{OM}(x,t) \ = F_{RP}(x,t) + F_{PT}(x,t) \quad (1)$$

While both force terms are a function of both the mirror displacement $x$ and time $t$, they represent delayed responses to the mirror position, and thus can be written as $F_{RP}(x,t) = F_{RP}\left(x\big(t-\tau_c(x)\big)\right)$ and $F_{PT}(x,t) = F_{PT}\big(x(t-\tau_c(x)-\tau_{th})\big)$. The delay $\tau_c$ is the retardation between a change in $x$ and change in radiation pressure force $F_{\mathrm{RP}}$, and is based upon the effective photon lifetime of the active cavity. The photothermal force is also delayed by $\tau_{th}$, the thermal response time constant of the HCG. Compared to the mechanical oscillation period (100 ns ~ 10 μs), $\tau_c$ is much smaller, on the orders of sub-nanosecond. While this delay can be approximated as small compared to the oscillation period, it is necessary to explain the anti-damping effect of optomechanical forces. If the thermal transient time of the HCG were comparable to the oscillation period, a lowpass term would be added to its delay expression. With Taylor expansion, both $F$ can be written as $F(x,t) = F\big(x(t) - \tau(t)\dot{x}(t)\big) = F(x) - \frac{dF}{dx} \cdot \tau(t)\dot{x}(t)$. Equation (1) can be re-arranged as:

$$m\ddot{x}(t) + \left[b + \frac{dF_{RP}}{dx} \cdot \tau_c \ + \ \frac{dF_{PT}}{dx} \cdot (\tau_c + \tau_{th})\right] \dot{x}(t) + [m\omega_o^2 x(t) - F_{OM}(x)] = 0 \qquad (2)$$

Equation (2) clearly reveals two functionalities of the optomechanical forces. Firstly, it changes the oscillation frequency, which can be described as the optical spring effect[29]. Secondly, an optomechanical damping term of form $\frac{dF}{dx} \cdot \tau(x)$ is added onto the original mechanical damping $b$ for each one of the optomechanical forces. With this in mind, the above equation can be rewritten

$$\ddot{x}(t) + [b + b_{RP} \ + \ b_{PT}]\dot{x}(t) + [m\omega_o^2 x(t) - F_{OM}(x)] = 0 \qquad (3)$$

where $b_{RP}$ and $b_{PT}$ represent radiation pressure and photothermal damping. When $b_{RP} + b_{PT} < 0$, optomechanical forces act as a mechanical gain, and if this anti-damping cancels the mechanical damping, regenerative oscillation can occur. Analysis of the laser rate equations can provide $\tau_c(x)$ and $\frac{dF_{RP}}{dx}$, and while $\frac{dF_{PT}}{dx}$ and $\tau_{th}$ additionally require measurements or simulation of the device's thermal expansion under photoabsorption. Thus, while photothermal forces are significantly smaller than radiation pressure in this high-reflectivity and low-absorption HCG mirror, the large thermal delay $\tau_{th}$ makes the net optomechanical gain, and thus power transfer, of photothermal effects comparable to radiation pressure in this device.

Compared to passive cavities, active optomechanical cavities offer optomechanical gain $b_{RP} + b_{PT} < 0$ over a wider position range, sustaining large-amplitude oscillation. Figure 5d-e presents a comparison between active-cavity and passive-cavity radiation-pressure optomechanics based on the locally linearized responses of the radiation pressure force to changes in mirror movement. This compares $\tau_c(x)$ and $b_{RP}$ for an active optomechanical cavity and a comparable passive optomechanical cavity, with the cold cavity optical Q and the output power of the laser optomechanics being the same as the input laser power of the passive cavity optomechanics. In this comparison, laser optomechanics shows a greatly enhanced $\tau_c(x)$ (>10X) in a much broader wavelength range (>200X) than passive-cavity optomechanics. This is attributed to the long carrier life time of the gain medium, due to large reservoir of carriers staying in the excited states. This serves as the buffer for the circulating photons inside the cavity, and effectively increases the photon lifetime. In the passive cavity, on the other hand, there is no such extra buffer, and the photon lifetime is solely determined by the passive cavity. Thus a very high-Q cavity is usually required. Another unique property of the laser optomechanics is that the high Q as well as the large optical circulating power can be maintained over a large mirror displacement, as the lasing wavelength is locked with the cavity. By contrast, in passive optomechanics, the displacement of the oscillator in a high

Q cavity is self-limited. This explains why the mirror displacement in laser optomechanics can be a few orders of magnitude larger than that in the passive optomechanics. In the regenerative oscillation regime, the oscillation amplitude is ultimately determined by the nonlinearity and the gain spectrum of the laser. There is also a great flexibility in designing the wavelength dependence of the mirror reflectivity and gain in laser optomechanics, which is key to further exploration of the optomechanical dynamics. Furthermore, since the mechanical damping factor $b$ should be as small as possible for optomechanical dynamics, the HCG VCSEL provides a unique advantage in this regard due to its two-order of magnitude lighter weight compared to conventional tunable MEMS-VCSELs[30]. Lastly, laser optomechanics with ultra-light HCG mirrors enables optomechanical coupling through both radiation pressure and photothermal forces, as shown in this work, offering large power transfer to the mechanical mode and an additional degree of configurability in designing optomechanical structures. For example, the use of bi-material structures or higher-absorptive materials in the HCG and its supports could further enhance photothermal effects above radiation pressure. The exact ratio of power transfer from light due to photothermal and radiation-pressure forces in the observed self-oscillation remains a subject for future work.

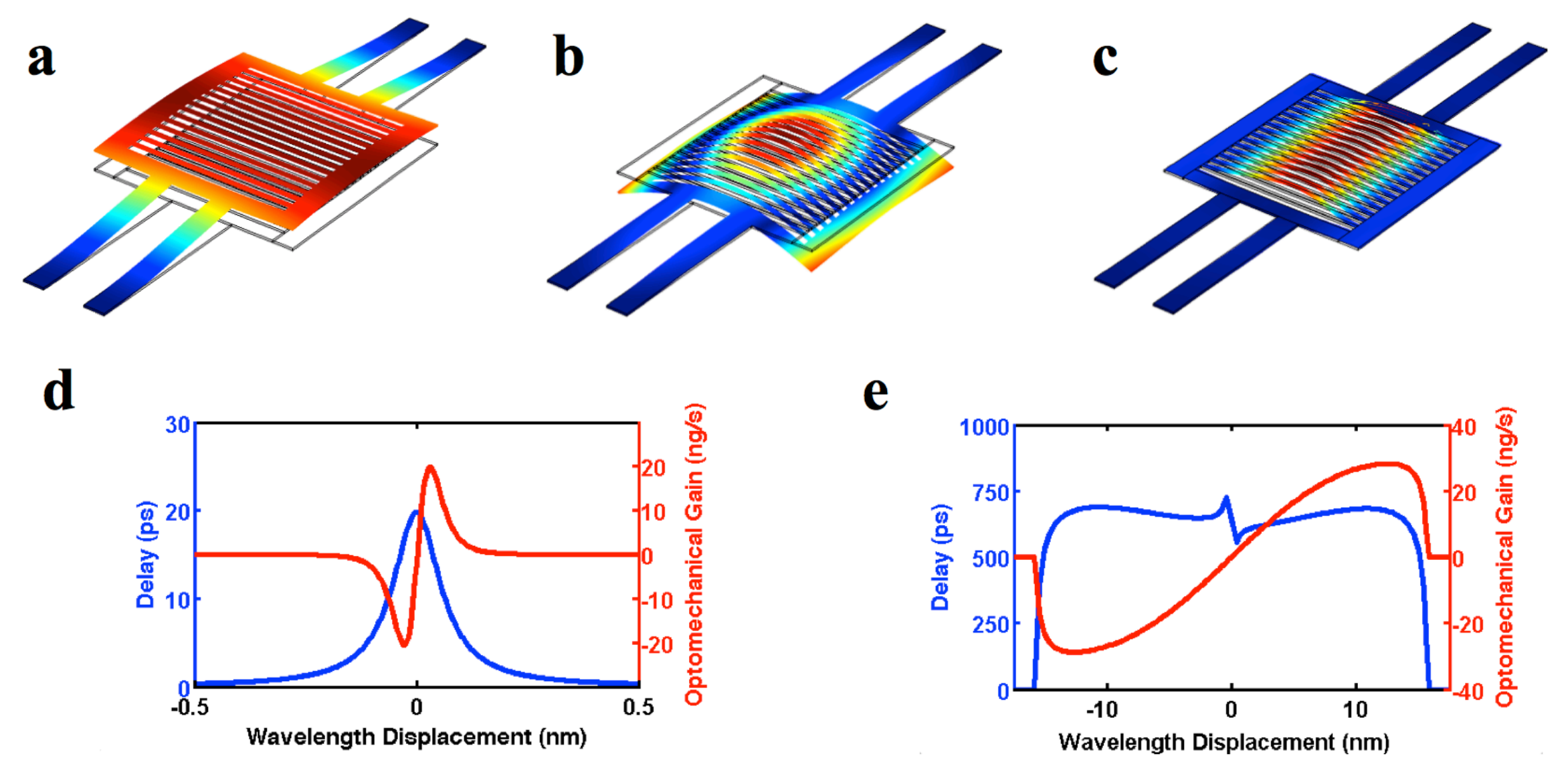

**Figure 5 | Mechanical modes of the HCG, and the analysis of the dynamical back action. a-c,** The mechanical modes of the HCG. **a** shows the fundamental mode, and **b-c** shows examples of the high order modes, where the HCG frame and bars are oscillating with the static bridges in **b**, and only the HCG bars are oscillating with the static frame and bridges in **c**. **d,** Optical delay $\tau_c(x)$ and optomechanical gain $-\frac{dF_{RP}}{dx}\cdot\tau_c(x)$ versus wavelength displacement for a radiation-pressure passive optomechanical cavity. **e,** Optical delay $\tau_c(x)$ and optomechanical gain $-\frac{dF_{RP}}{dx}\cdot\tau_c(x)$ versus wavelength displacement for an equivalent radiation-pressure laser optomechanical cavity.

To conclude, we have demonstrated a new laser optomechanics phenomenon using a VCSEL with an ultralight HCG mirror. A strong power transfer between photons and a mechanical resonator is achieved by collocating a laser cavity with a mechanical oscillator. We show that four key criteria are necessary to achieve high-amplitude self-oscillation: (1) an ultra-low-mass mechanical oscillator, (2) a strong wavelength dependence of the mirror reflectivity, (3) a high mechanical quality factor, and (4) a sufficient delay from the laser rate equations due to long photon lifetime (optical Q) and long carrier lifetime in the active region. Such a laser optomechanical system will extend the functionality of tunable lasers by creating a high-speed wide-range wavelength-swept source for applications in optical coherence tomography (OCT), light detection and ranging (LIDAR) and 3D cameras, and will establish a platform to further study laser optomechanics.

## METHODS SUMMARY

The 1550 nm HCG VCSEL is fabricated on an InP epitaxial wafer with AlGaInAs as the quantum wells. The HCG is defined by electron beam lithography followed by a wet chemical etch. The current aperture is defined by proton implantation. The HCG is released by selective wet chemical etch. See Ref. 24 for a detailed fabrication flow.

The mechanical quality factor of the MEMS HCG is measured independently from the optomechanical effects with a LDV system (Polytec OFV-3001). The HCG VCSEL is placed under vacuum to reduce the squeeze-film mechanical damping of the HCG in air. This technique resolves the velocity of a moving MEMS device using the Doppler frequency shift of the reflected beam from an external laser beam onto the device. Since the laser optomechanical cavity is not pumped and the external laser wavelength is far from the cavity resonance, the measurement will not be perturbed by the device's optomechanical effects.

To characterize the optomechanical properties of the HCG VCSEL, the VCSEL is placed in a vacuum chamber, and a multimode cleaved fiber is used to couple the VCSEL output light into the fiber. The optical power is then split into three paths. One path is connected to a photodetector with 1.2 GHz 3-dB bandwidth. Another path is connected to a monochromator followed by a photodetector with the same bandwidth. The outputs of the two photodetectors are connected to power amplifiers with 250 MHz 3-dB bandwidth, followed by a real-time oscilloscope, for time-resolved optical power and lasing wavelength measurement. The pass-band of the monochromator is scanned across the wavelength swept range of the VCSEL, and the time-resolved lasing wavelength can be extracted from the series of traces obtained in the oscilloscope. Alternatively, the output of the amplifier can be connected to a RF spectrum analyzer to identify the mechanical modes and their properties. The third path of the split optical light is connected to an optical spectrum analyzer to characterize its average optical spectrum.

**Acknowledgements** We are grateful to M. C. Wu for useful discussions and for providing the LDV system for measurements, and F. Ogletree for his support on the scanning electron microscopy on the device. We also thank Y. Wang and B. Behroozpour for assistance performing the LDV measurements. C.C.H. acknowledges financial support from a Department of Defense National Security Science and Engineering Faculty Fellowship. Work at the Molecular Foundry was supported by the Office of Science, Office of Basic Energy Sciences.

**Author Contributions** C.C.H. proposed and guided the overall project. W.Y., S.A.G., and C.C.H. designed the experiments. W.Y. and S.A.G performed the experiment, simulation and analysis. K.W.N and W.Y. performed the scanning electron microscopy. C.C designed the device. Y.R. fabricated the device. W.Y., S.A.G., and C.C.H. composed the manuscript. All authors discussed the results and commented on the manuscript.

**Author Information** Correspondence and requests for materials should be addressed to C.C.H. (cch@berkeley.edu).